\documentstyle[twoside]{article}

\catcode`\@=11
\long\def\@makefntext#1{
\protect\noindent \hbox to 3.2pt {\hskip-.9pt  
$^{{\eightrm\@thefnmark}}$\hfil}#1\hfill}  

\def\@makefnmark{\hbox to 0pt{$^{\@thefnmark}$\hss}} 
 
\def\ps@myheadings{\let\@mkboth\@gobbletwo
\def\@oddhead{\hbox{}
\rightmark\hfil\eightrm\thepage}   
\def\@oddfoot{}\def\@evenhead{\eightrm\thepage\hfil
\leftmark\hbox{}}\def\@evenfoot{}
\def\sectionmark##1{}\def\subsectionmark##1{}}

\oddsidemargin=\evensidemargin
\newcounter{sectionc}\newcounter{subsectionc}\newcounter{subsubsectionc}
\renewcommand{\section}[1] {\vspace{12pt}\addtocounter{sectionc}{1} 
\setcounter{subsectionc}{0}\setcounter{subsubsectionc}{0}\noindent 
 {\tenbf\thesectionc. #1}\par\vspace{5pt}}
\renewcommand{\subsection}[1] {\vspace{12pt}\addtocounter{subsectionc}{1} 
 \setcounter{subsubsectionc}{0}\noindent 
 {\bf\thesectionc.\thesubsectionc. {\kern1pt \bfit #1}}\par\vspace{5pt}}
\renewcommand{\subsubsection}[1]
{\vspace{12pt}\addtocounter{subsubsectionc}{1}
 \noindent{\tenrm\thesectionc.\thesubsectionc.\thesubsubsectionc.
 {\kern1pt \tenit #1}}\par\vspace{5pt}}
\newcommand{\nonumsection}[1] {\vspace{12pt}\noindent{\tenbf #1}
 \par\vspace{5pt}}

\newcounter{appendixc}
\newcounter{subappendixc}[appendixc]
\newcounter{subsubappendixc}[subappendixc]
\renewcommand{\thesubappendixc}{\Alph{appendixc}.\arabic{subappendixc}}
\renewcommand{\thesubsubappendixc}
 {\Alph{appendixc}.\arabic{subappendixc}.\arabic{subsubappendixc}}

\renewcommand{\appendix}[1] {\vspace{12pt}
        \refstepcounter{appendixc}
        \setcounter{figure}{0}
        \setcounter{table}{0}
        \setcounter{lemma}{0}
        \setcounter{theorem}{0}
        \setcounter{corollary}{0}
        \setcounter{definition}{0}
        \setcounter{equation}{0}
        \renewcommand{\thefigure}{\Alph{appendixc}.\arabic{figure}}
        \renewcommand{\thetable}{\Alph{appendixc}.\arabic{table}}
        \renewcommand{\theappendixc}{\Alph{appendixc}}
        \renewcommand{\thelemma}{\Alph{appendixc}.\arabic{lemma}}
        \renewcommand{\thetheorem}{\Alph{appendixc}.\arabic{theorem}}
        \renewcommand{\thedefinition}{\Alph{appendixc}.\arabic{definition}}
        \renewcommand{\thecorollary}{\Alph{appendixc}.\arabic{corollary}}
        \renewcommand{\theequation}{\Alph{appendixc}.\arabic{equation}}
        \noindent{\tenbf Appendix#1}\par\vspace{5pt}}
\newcommand{\subappendix}[1] {\vspace{12pt}
        \refstepcounter{subappendixc}
        \noindent{\bf Appendix \thesubappendixc. {\kern1pt \bfit #1}}
 \par\vspace{5pt}}
\newcommand{\subsubappendix}[1] {\vspace{12pt}
        \refstepcounter{subsubappendixc}
        \noindent{\rm Appendix \thesubsubappendixc. {\kern1pt \tenit #1}}
 \par\vspace{5pt}}

\newcommand{\textlineskip}{\baselineskip=13pt}
\newcommand{\smalllineskip}{\baselineskip=10pt}

\def\eightcirc{
\begin{picture}(0,0)
\put(4.4,1.8){\circle{6.5}}
\end{picture}}
\def\eightcopyright{\eightcirc\kern2.7pt\hbox{\eightrm c}} 

\newcommand{\copyrightheading}[1]
 {\vspace*{-2.5cm}\smalllineskip{\flushleft
 {\footnotesize Mathematical Models and Methods in Applied Sciences #1}\\
 {\footnotesize $\eightcopyright$\, World Scientific Publishing
  Company}\\
  }}

\def\abstracts#1#2#3{{
 \centering{\begin{minipage}{4.5in}\baselineskip=10pt\footnotesize
 \parindent=0pt #1\par 
 \parindent=15pt #2\par
 \parindent=15pt #3
 \end{minipage}}\par}} 

\renewenvironment{thebibliography}[1]
 {\frenchspacing
  \ninerm\baselineskip=11pt
  \begin{list}{\arabic{enumi}.}
        {\usecounter{enumi}\setlength{\parsep}{0pt}     
  \setlength{\leftmargin 12.7pt}{\rightmargin 0pt} 
         \setlength{\itemsep}{0pt} \settowidth
 {\labelwidth}{#1.}\sloppy}}{\end{list}}

\newcounter{itemlistc}
\newcounter{romanlistc}
\newcounter{alphlistc}
\newcounter{arabiclistc}

\newcommand{\fcaption}[1]{
        \refstepcounter{figure}
        \setbox\@tempboxa = \hbox{\footnotesize Fig.~\thefigure. #1}
        \ifdim \wd\@tempboxa > 5in
           {\begin{center}
        \parbox{5in}{\footnotesize\smalllineskip Fig.~\thefigure. #1}
            \end{center}}
        \else
             {\begin{center}
             {\footnotesize Fig.~\thefigure. #1}
              \end{center}}
        \fi}

\newcommand{\tcaption}[1]{
        \refstepcounter{table}
        \setbox\@tempboxa = \hbox{\footnotesize Table~\thetable. #1}
        \ifdim \wd\@tempboxa > 5in
           {\begin{center}
        \parbox{5in}{\footnotesize\smalllineskip Table~\thetable. #1}
            \end{center}}
        \else
             {\begin{center}
             {\footnotesize Table~\thetable. #1}
              \end{center}}
        \fi}

\def\@citex[#1]#2{\if@filesw\immediate\write\@auxout
 {\string\citation{#2}}\fi
\def\@citea{}\@cite{\@for\@citeb:=#2\do
 {\@citea\def\@citea{,}\@ifundefined
 {b@\@citeb}{{\bf ?}\@warning
 {Citation `\@citeb' on page \thepage \space undefined}}
 {\csname b@\@citeb\endcsname}}}{#1}}

\newif\if@cghi
\def\cite{\@cghitrue\@ifnextchar [{\@tempswatrue
 \@citex}{\@tempswafalse\@citex[]}}
\def\citelow{\@cghifalse\@ifnextchar [{\@tempswatrue
 \@citex}{\@tempswafalse\@citex[]}}
\def\@cite#1#2{{$\null^{#1}$\if@tempswa\typeout
 {IJCGA warning: optional citation argument 
 ignored: `#2'} \fi}}

\def\pmb#1{\setbox0=\hbox{#1}
 \kern-.025em\copy0\kern-\wd0
 \kern.05em\copy0\kern-\wd0
 \kern-.025em\raise.0433em\box0}

\def\fnt#1#2{\footnotetext{\kern-.3em
 {$^{\mbox{\scriptsize #1}}$}{#2}}}

\def\fpage#1{\begingroup
\voffset=.3in
\thispagestyle{empty}\begin{table}[b]\centerline{\footnotesize #1}
 \end{table}\endgroup}

\def\runninghead#1#2{\pagestyle{myheadings}
\markboth{{\protect\footnotesize\it{\quad #1}}\hfill}
{\hfill{\protect\footnotesize\it{#2\quad}}}}
\font\tenrm=cmr10
\font\tenit=cmti10 
\font\tenbf=cmbx10
\font\bfit=cmbxti10 at 10pt
\font\ninerm=cmr9

\font\eightrm=cmr8

\def\qed{\hbox{${\vcenter{\vbox{   
   \hrule height 0.4pt\hbox{\vrule width 0.4pt height 6pt
   \kern5pt\vrule width 0.4pt}\hrule height 0.4pt}}}$}}

\def\theequation{\thesectionc.\arabic{equation}} 

\begin{document}

\runninghead{Presented at Dublin Finance Conference (1999)}
{}

\normalsize\textlineskip
\thispagestyle{empty}
\setcounter{page}{1}

\copyrightheading{}   

\vspace*{0.88truein}

\fpage{1}
\centerline{\bf TRADER DYNAMICS IN A MODEL MARKET}
\vspace*{0.035truein}
\centerline{\bf }
\vspace*{0.37truein}
\centerline{\footnotesize NEIL F. JOHNSON  
 and MICHAEL HART}
\vspace*{0.015truein}
\centerline{\footnotesize\it Physics Department, Oxford University}
\baselineskip=10pt
\centerline{\footnotesize\it Parks Road, Oxford, OX1 3PU, U.K.}
\baselineskip=10pt
\centerline{\footnotesize\it n.johnson@physics.ox.ac.uk}
\vspace*{10pt}
\centerline{\footnotesize PAK MING HUI}
\vspace*{0.015truein}
\centerline{\footnotesize\it Department of Physics, The Chinese University of
Hong Kong}
\baselineskip=10pt
\centerline{\footnotesize\it Shatin, New Territories, Hong Kong}
\baselineskip=10pt
\centerline{\footnotesize\it pmhui@phy.cuhk.edu.hk}
\vspace*{10pt}
\centerline{\footnotesize DAFANG ZHENG}
\vspace*{0.015truein}
\centerline{\footnotesize\it Department of Applied Physics, South China
University of Technology}
\baselineskip=10pt
\centerline{\footnotesize\it Guangzhou 510641, People's Republic of China}
\baselineskip=10pt
\centerline{\footnotesize\it phdzheng@scut.edu.cn}
\vspace*{0.225truein}

\vspace*{0.4truein}
\abstracts{We explore various extensions of 
Challet and Zhang's Minority Game in an
attempt to gain insight into the dynamics underlying financial markets.
First we consider a
heterogeneous population where individual traders employ differing `time
horizons' when 
making predictions based on historical data. The resulting average winnings
per trader is a
highly non-linear function of the population's composition. Second, we
introduce a
threshold confidence level among traders below which they will not trade.
This
can give rise to large fluctuations in the `volume' of market participants
and the
resulting market `price'.  }{}{}



\vspace*{1pt}\textlineskip 
\section{Introduction} 
\vspace*{-0.5pt}
\noindent
Two obvious practical goals in the study of financial markets are
to understand how arbitrage opportunities might arise and consequently be
exploited, and
to understand quantitatively  the origin of price fluctuations.
The Minority
Game\cite{challet1,challet2,savit,johnson2,rodgers,decara,cavagna}
represents a fascinating toy-model of a complex adaptive system in which 
individual members
(e.g., traders) repeatedly compete to be in a minority. The Minority Game
offers a simple
paradigm for the decision dynamics underlying financial markets: if for
example there
are more buyers than sellers at a given moment, prices are pushed up and
hence it would
be better for a trader to be in the minority group of sellers.

Here we explore two extensions of the Minority Game which seem relevant for
real
markets. First we consider the
performance of a heterogeneous population of traders who differ in the 
`time horizon' employed when making buy/sell decisions based on past market
data. We find
that the average winnings per trader is a highly non-linear function of the
population's
composition. Second, we introduce a threshold confidence level among traders
below which
they will not trade. We find that this feature can give rise to large
variations in the
`volume' of market participants and the resulting market `price'. 


\section{Basic Minority Game}
\noindent
The basic Minority Game
\cite{challet1,challet2,savit,johnson2,rodgers,decara,cavagna}  consists 
of a repeated game with an
odd number of traders $N$ who must choose independently whether to be in
room 0 (e.g. buy)
or room 1 (e.g. sell).   The winners are those in the room with fewer
traders, i.e. the sellers win if there is an excess of buyers. The output is
a single
binary digit, 0 or 1, representing the winning decision for each time step. 
This output
is made available to all traders, and is the only
information they can use to make decisions in subsequent
turns.   The memory $m$ is the length of the recent history
bit-string that a trader uses when making its next decision
\cite{challet1,challet2,savit,johnson2,rodgers,decara}. In the market
context, $m$ can be thought of as a `time horizon' over which a given trader
considers the
past history to be relevant when making a prediction as to the direction of
the next market
movement. 

The traders randomly pick
$s$ strategies at the beginning of the game, with repetitions
allowed. After each turn, the trader assigns one (virtual)
point to each of his strategies which would have predicted the
correct outcome. In addition the trader gets awarded one (real)
point if he is successful. At each turn of the game, the trader
uses the most successful strategy (i.e., most virtual points) 
from his bag of
$s$ strategies.   
The strategy-space forms a
$2^m$-dimensional hypercube for memory $m$ with strategies at
the $2^{2^m}$ vertices\cite{challet1,challet2}. If the size of the
strategy space is small  compared to the total number of
traders $N$ (i.e., $2\cdot 2^m<<Ns$) 
many traders may hold the
highest-scoring strategy at any given turn and hence make the
same decision\cite{challet1,challet2,savit,johnson2}.
This leads to a large standard deviation in the
winning room and hence a relatively low number of total points
awarded
\cite{challet1,challet2,savit,johnson2,rodgers,decara}.  Such
crowd-effects are a strategy-space phenomenon and have been
shown\cite{johnson2} to quantitatively explain the
fluctuations for the pure
population as a function of $m$ and
$s$.   

\section{Mixed population of traders}
\noindent
Consider a population containing
$N_{m_{1}}$ traders with memory $m_{1}$, and $N_{m_{2}}=N-N_{m_{1}}$
traders with
memory $m_{2}$.  We define the average points per trader
per turn, $W$, to be the total number of points awarded in
that turn divided by the total number of traders.   For small
$m$, $W$ is substantially less than 0.5 due to the
crowd-effects mentioned above. The maximum possible 
$W$ would correspond to the number of winners remaining at
$(N-1)/2$. Therefore 
$W$ is always less than or equal to $(N-1)/2N$, hence
$W<0.5$.  Note that an external (i.e. non-participating)
gambler using a coin-toss to predict the winning decision, would
have a
$50\%$ success rate since he would not suffer from this
intrinsic crowding in strategy-space.
The history-space forms an 
$m$-dimensional hypercube whose $2^m$ vertices correspond to
all  possible recent history bit-strings of length $m$. 
For a {\it pure} population of traders with the same
memory $m$, where $2\cdot 2^m<<Ns$, there is information left in
the history time-series\cite{savit}: however this information is hidden in
bit-strings of length greater than
$m$ and hence is not accessible to these traders.
For large
$m$, there is information left in bit-strings of any length:
however the traders have insufficient strategies to further
exploit this information\cite{savit}.  

\begin{figure}[htbp]
\vspace*{13pt}
\centerline{\vbox{\hrule width 5cm height0.001pt}}
\vspace*{1.6truein}  
\centerline{\vbox{\hrule width 5cm height0.001pt}}
\vspace*{13pt}
\fcaption{Numerical results for average winnings
per trader per turn $W$:
traders with memory $m=3$ possess $s_{1}=2,3,\cdots,7$
strategies (each data set corresponds to a different $s_{1}$ value)
whereas
traders with memory
$m=6$ have
$s_{2}=7$ strategies for each data set. 
}
\end{figure}

Figure 1 shows the numerical results for the
average points per trader per turn
$W$ in a mixed population of 
$N=101$ traders, with memory $m=3$ or $6$. 
The number of
$m=3$ traders
$N_3$ is shown on the $x$-axis, hence the number of $m=6$
traders is given by $N_6=101-N_3$. 
The $m=3$ traders possess $s_{1}=2,3,\cdots,7$
strategies (each data set corresponds to a different $s_{1}$) 
while the $m=6$ traders possess
$s_{2}=7$ strategies. 
The data were collected in the limit of long times, and averaged over many
runs. 
We observe from Fig. 1 that the average winnings $W$ can show a maximum at
finite mixing.

\begin{figure}[htbp]
\vspace*{13pt}
\centerline{\vbox{\hrule width 5cm height0.001pt}}
\vspace*{1.6truein}  
\centerline{\vbox{\hrule width 5cm height0.001pt}}
\vspace*{13pt}
\fcaption{Numerical results for average winnings per trader per turn $W$: 
traders with memory $m=3$ possess $s_1=7$
strategies for each data set whereas
traders with memory
$m=6$ have $s_2=2,3,\cdots,7$ strategies
(each data set corresponds to a different
$s_2$ value) . 
}
\end{figure}

Figure 2 shows results for the opposite case where the $m=3$ traders each
possess $s_{1}=7$
strategies while the $m=6$ traders possess
$s_{2}=2,3,\cdots,7$ strategies. 
Note that simulations in which traders
are fed a random (as opposed to real) history\cite{cavagna}
do not reproduce the numerical results of Figs. 1 and 2. This is
essentially because
the $m=6$ traders in the real-history game have the opportunity to
exploit correlations in the real-history time-series left by
the $m=3$ traders (see also Refs. [1],[2] and [8]). In other words, the
long-memory (i.e.,
$m=6$) traders can identify and exploit arbitrage opportunities that are
inaccessible to the short-memory (i.e., $m=3$) traders.

Any fluctuation in the number of winners away from
$(N-1)/2=50$ implies wasteage of total points\cite{challet1,challet2}.
Hence $W \sim 0.5-\frac{\sigma}{N}$, where $\sigma$ is the
standard deviation of the number of traders making a given decision, say
`buy'.
In order to develop an analytic theory, we assume that the
corresponding
$\sigma$ can be obtained by adding separately the contributions to
the variance from the $m=3$ traders and the
$m=6$ traders.  This amounts to assuming 
that the  system has managed to
remove any internal frustration between traders with different
memories, and the two groups of traders behave independently. Hence
$\sigma^2 \sim
\sigma_3^2 + \sigma_6^2$, where $\sigma_3$ ($\sigma_6$)
is the variance due to the $m=3$ ($6$) traders.
Following Ref. [4],  and
defining the concentration of $m=3$ traders as
$x=N_3/N$, we obtain $\sigma_3 \sim C_3 x N$ and
$\sigma_6 \sim C_6 (1-x) N$ where $C_3$ and $C_6$
are given by the general
expression 
\begin{equation}
C_m=\frac{1}{2} \bigg[ \sum_r \bigg(\bigg[1 -
\frac{(r-1)}{2\cdot 2^m}\bigg]^s -
\bigg[1-\frac{r}{2\cdot 2^m}\bigg]^s\bigg)^2 \bigg]^\frac{1}{2}
\ .
\end{equation}
The summation is over weakly correlated groups: each group $r$ comprises a
crowd of
like-minded (i.e., correlated) traders who use the same strategy, and a
corresponding 
anticrowd of opposite-minded traders who use the anticorrelated
strategy\cite{johnson2}. 
The short-memory (i.e., $m=3$) sub-population of traders
tends to lie in the crowded regime, hence we only need to include $r=1$ in
the
summation to obtain
$C_3$.  The long-memory (i.e., $m=6$)
sub-population of traders will tend to lie in the
crowded regime if the number $s_{2}$ of
strategies they hold is large,
but will form
crowd-anticrowd pairs if $s_2$ is small\cite{johnson2}.
To obtain $C_6$ we
sum up the terms from
$r=1$ to $r=101(1-x)$
if $\frac{2.2^6}{2}\geq 101(1-x)$, or to
$r=\frac{2.2^6}{2}$ if
$\frac{2.2^6}{2}<101(1-x)$.
The analytic expression for the average winnings
is given by 
\begin{equation}
W\sim 0.5 - [C_3^2 x^2 + C_6^2
(1-x)^2]^{1/2}\ .
\end{equation} 

\begin{figure}[htbp]
\vspace*{13pt}
\centerline{\vbox{\hrule width 5cm height0.001pt}}
\vspace*{1.6truein}  
\centerline{\vbox{\hrule width 5cm height0.001pt}}
\vspace*{13pt}
\fcaption{Analytic results corresponding to Figure 1. 
}
\end{figure}

\begin{figure}[htbp]
\vspace*{13pt}
\centerline{\vbox{\hrule width 5cm height0.001pt}}
\vspace*{1.6truein}  
\centerline{\vbox{\hrule width 5cm height0.001pt}}
\vspace*{13pt}
\fcaption{Analytic results corresponding to Figure 2.
}
\end{figure}

Figures 3 and 4 show the analytic results corresponding to the numerical
simulations of
Figs. 1 and 2 respectively. As can be seen the agreement is fairly good.
Intriguingly,
the theory has a tendency to underestimate the 
actual winnings suggesting that the
actual population is somehow exhibiting an additional degree of co-operation
(correlation). A
fuller theory of this mixed system thus requires the inclusion
of higher-order inter-trader correlations.

\section{Threshold confidence level among traders}
\noindent
In the Minority Game with either a homogeneous or heterogeneous population,
traders
must either buy or sell at every time-step. In a real market, however,
traders are likely to
wait on the sidelines until they are reasonably confident of winning at a
given
time-step. They will observe the market passively, mentally updating their
various strategies,
until their confidence overcomes some threshold value - then they will jump
in and trade.

We now attempt to incorporate this general behavior as follows.  Taking the
simplest
generalization, we assign a threshold confidence level $r_{\rm min}$ below
which a trader
will not trade. A trader's confidence level at a given time-step is
determined by
the success rate $r$ of his best-performing strategy over the last $T$
time-steps. 
Hence the
number of traders who actively trade $N_{\rm active}$  will fluctuate in
time - this feature
is reminiscent of the grand canonical ensemble in statistical 
mechanics\cite{challet1,challet2}. We will call $N_{\rm active}(t)$ the
`volume' of
active traders at time
$t$. Given that $N_{\rm active}(t)=N_{\rm buy}(t) + N_{\rm sell}(t)$, we can
form a
simple-minded `price' time-series $P(t)$ by setting 
\begin{equation}
P(t+1)=P(t) + [N_{\rm buy}(t) - N_{\rm sell}(t)]/D 
\end{equation}
where $D$ is a parameter characteristic of a particular market.
For simplicity we take $D$ to be
time-independent. Similar linear expressions
for the price $P(t)$ have been discussed
recently by other authors\cite{price}.

A real market is not a zero-sum game - the commission charged by the market
maker and/or
general transaction costs imply that the success rate will be less than
$50\%$, similar to
the Minority Game. Hence it is reasonable to expect that typically $r_{\rm
min}\geq 0.5$.
Interestingly, the fluctuations in $N_{\rm
active}$ are much larger at
$r_{\rm min}=0.5$ than for $r_{\rm min}\neq 0.5$:
in fact the resulting plot of fluctuations
as a function of $r_{\rm min}$
resembles a
$\lambda$-transition in statistical mechanics\cite{johnson3}.
Hence $r_{\rm min}=0.5$ seems to
represent a `critical' value, thereby
adding weight to the conjecture that real markets
may be positioned at some kind of critical point.

\begin{figure}[htbp]
\vspace*{13pt}
\centerline{\vbox{\hrule width 5cm height0.001pt}}
\vspace*{2.8truein}  
\centerline{\vbox{\hrule width 5cm height0.001pt}}
\vspace*{13pt}
\fcaption{A section of the ``price" and ``volume"
time-series generated by
minority game with
confidence threshold $r_{\rm min}=0.51$ and long trader memory ($m=6$).
$N=1001$ and $s=2$.
}
\end{figure}

\begin{figure}[htbp]
\vspace*{13pt}
\centerline{\vbox{\hrule width 5cm height0.001pt}}
\vspace*{2.8truein}  
\centerline{\vbox{\hrule width 5cm height0.001pt}}
\vspace*{13pt}
\fcaption{A section of the ``price" and ``volume"
time-series generated by
minority game with
confidence threshold $r_{\rm min}=0.51$ and short trader memory ($m=2$).
$N=1001$ and $s=2$.
}
\end{figure}

Figures 5 and 6 show the resulting ``price"
and ``volume" series for trader
populations with
long and short memories respectively. In the simulations
we take $T=500$, however the results are similar for
larger $T$.  For a population of the long-memory traders
(Fig. 5), the
volume is  non-zero and large jumps in the price do not occur at a single
time-step. Although
similar general patterns can be found in the
time-series for a population of short-memory
traders (Fig. 6)
the volume is often zero (the market becomes illiquid) and exhibits large
spikes -
corresponding large jumps in the price also arise (Fig. 6). Although this study is at
a preliminary
stage, we note in passing that large jumps in real stock prices can indeed be
accompanied by large
jumps in trading volume: see, for example, the behavior of Vodafone
stock\cite{charting} in November
1995, and Hanson stock\cite{charting} at the end of January 1996.  

\pagebreak

\section{Conclusion}
\noindent
In summary, we have discussed two generalizations of the basic Minority
Game. 
When the trader population is characterized by two different memories (`time
horizons') the
average winnings per trader can exceed that of a pure population. When
traders can opt not
to trade based on their confidence level, we find that large fluctuations
can arise in the
`volume' of market participants and the resulting market `price'.

\nonumsection{Acknowledgment}
\noindent
We thank D. Challet for discussions.

\nonumsection{References}

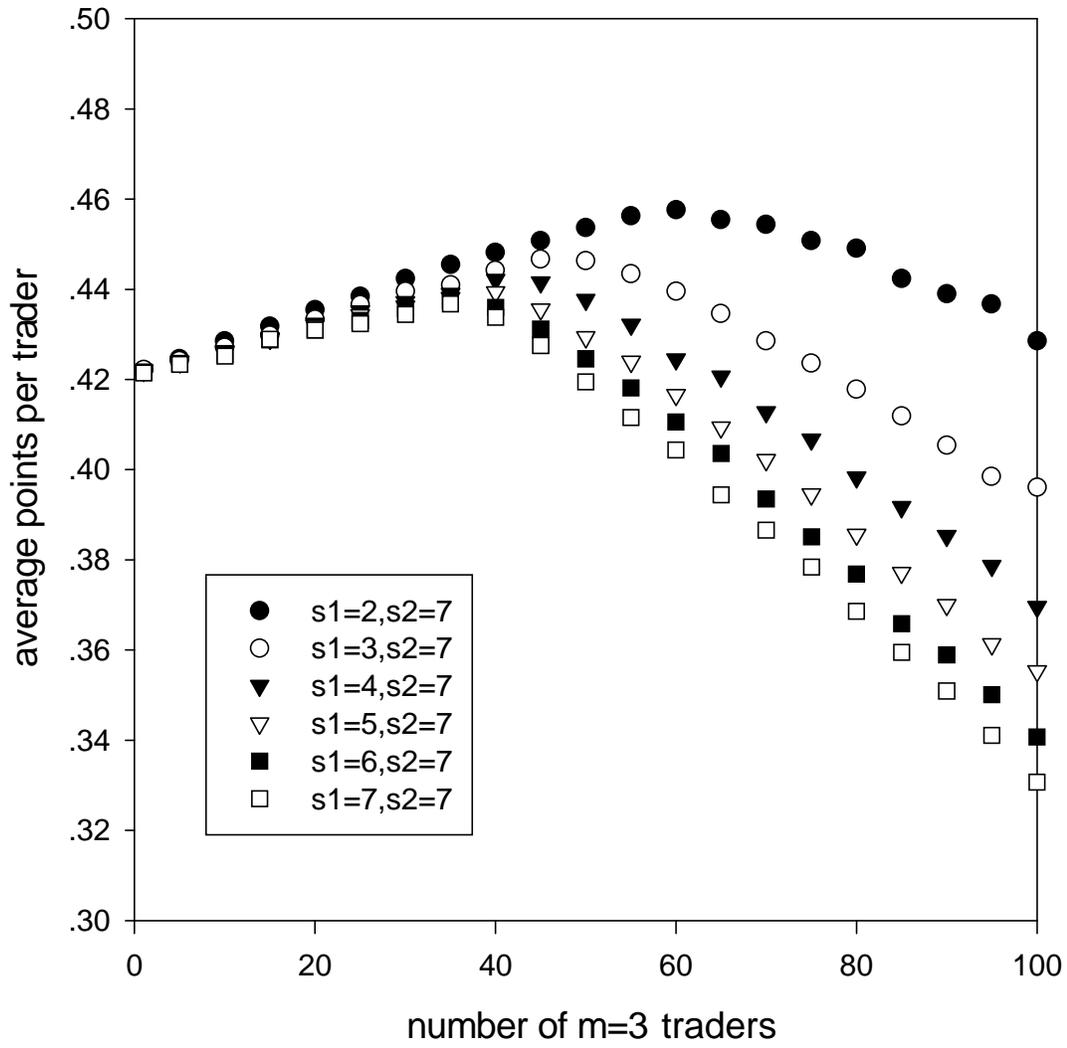
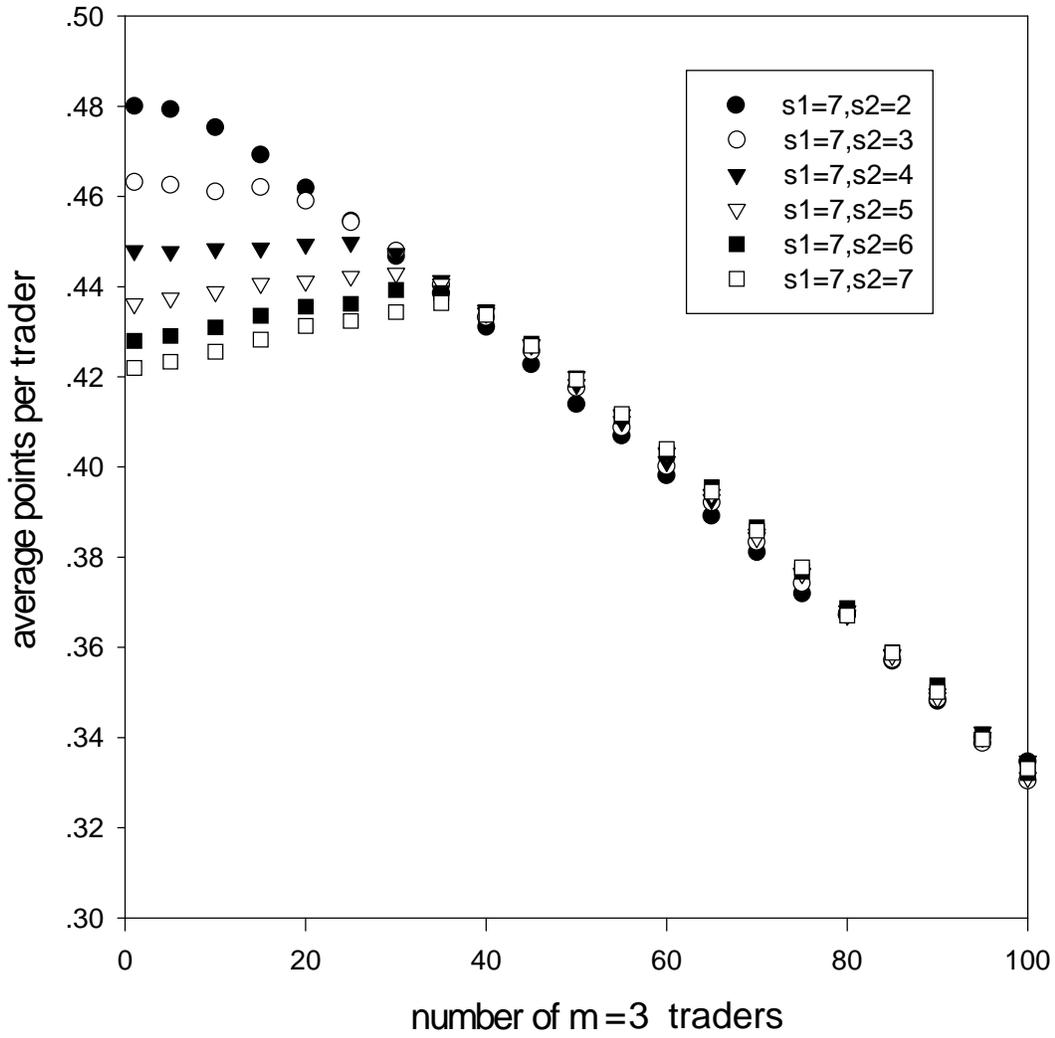
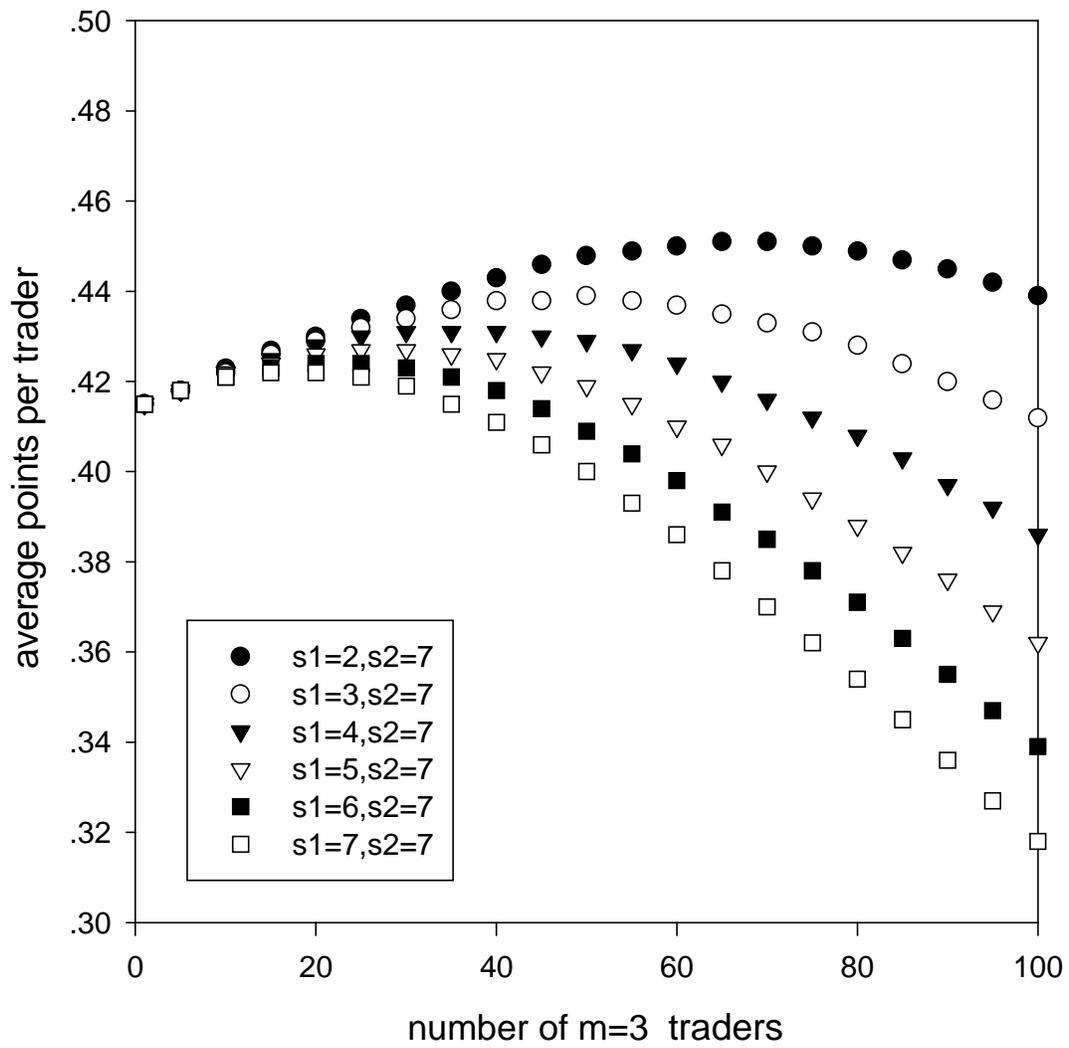
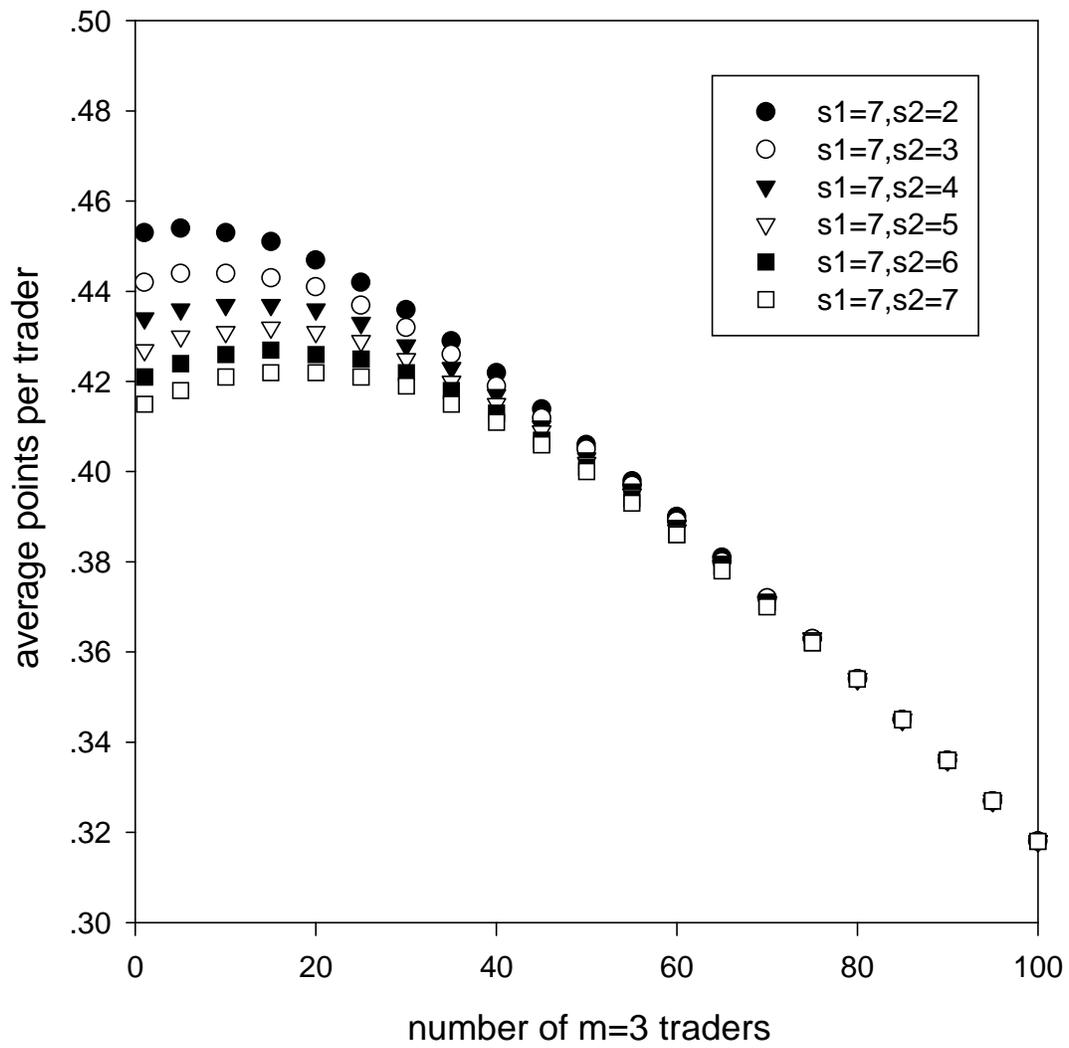
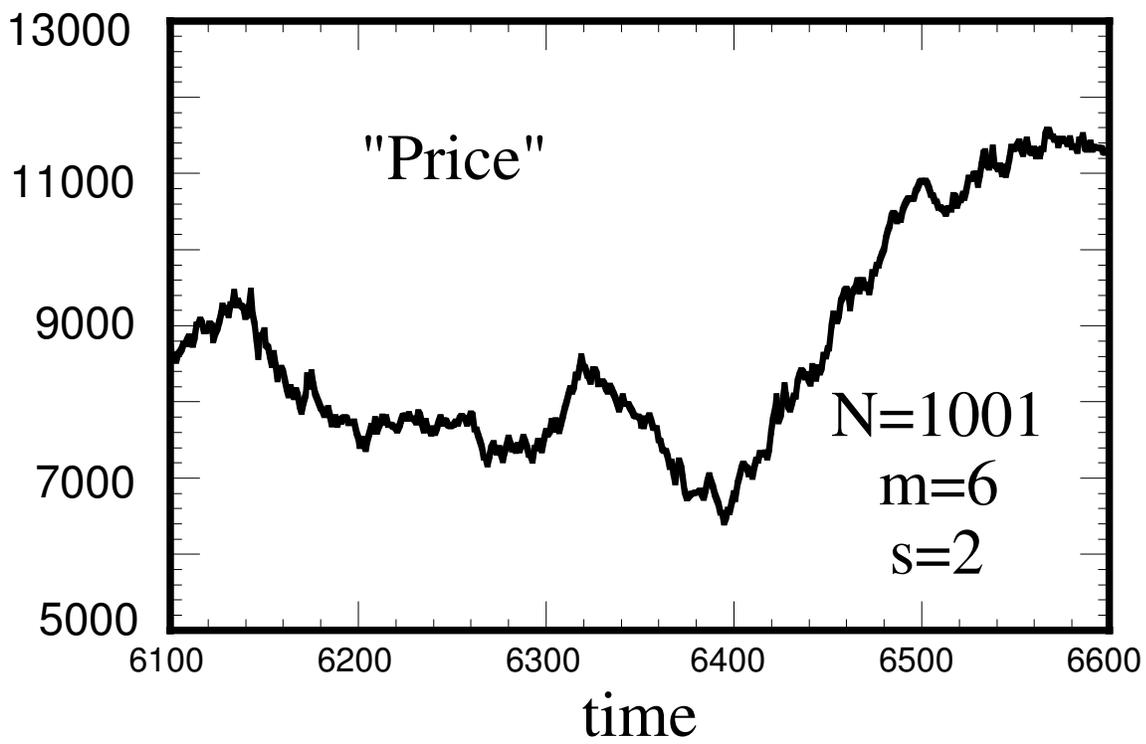
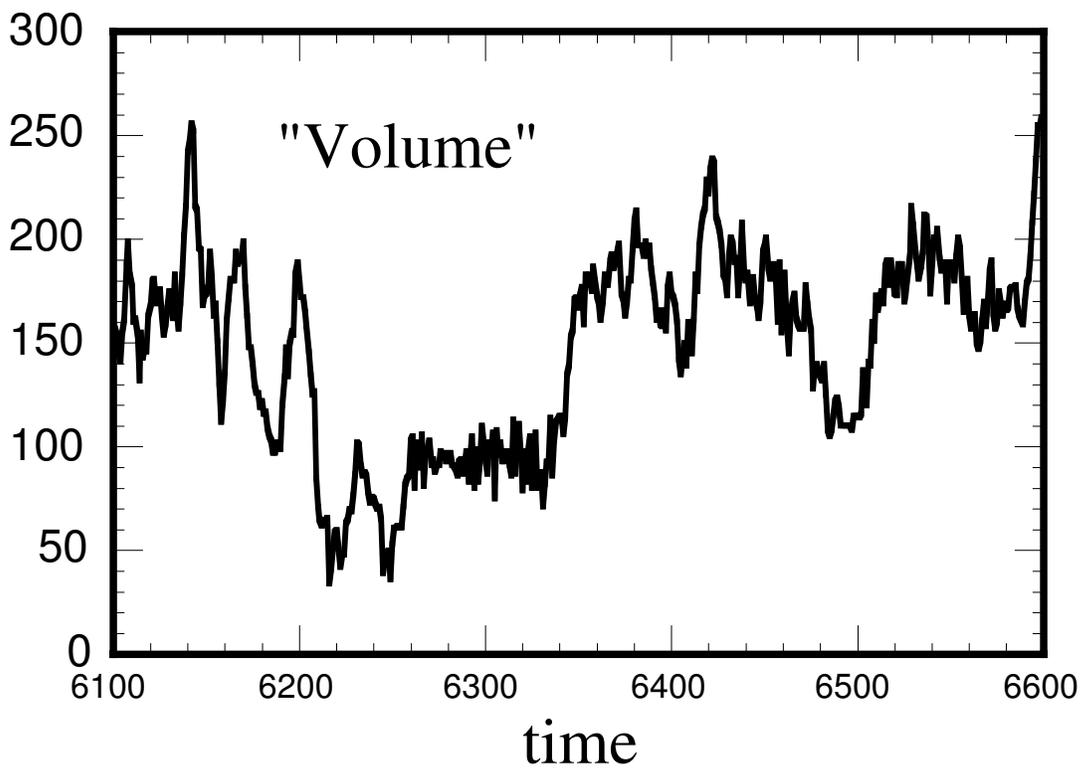
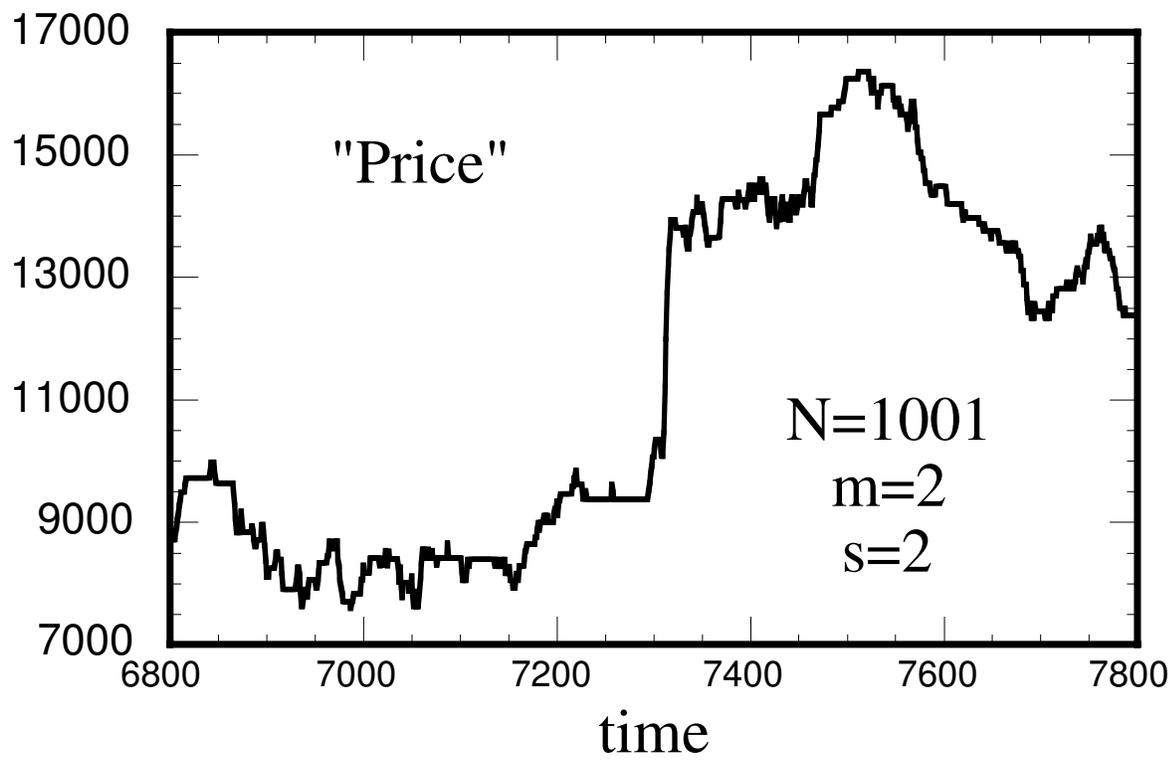
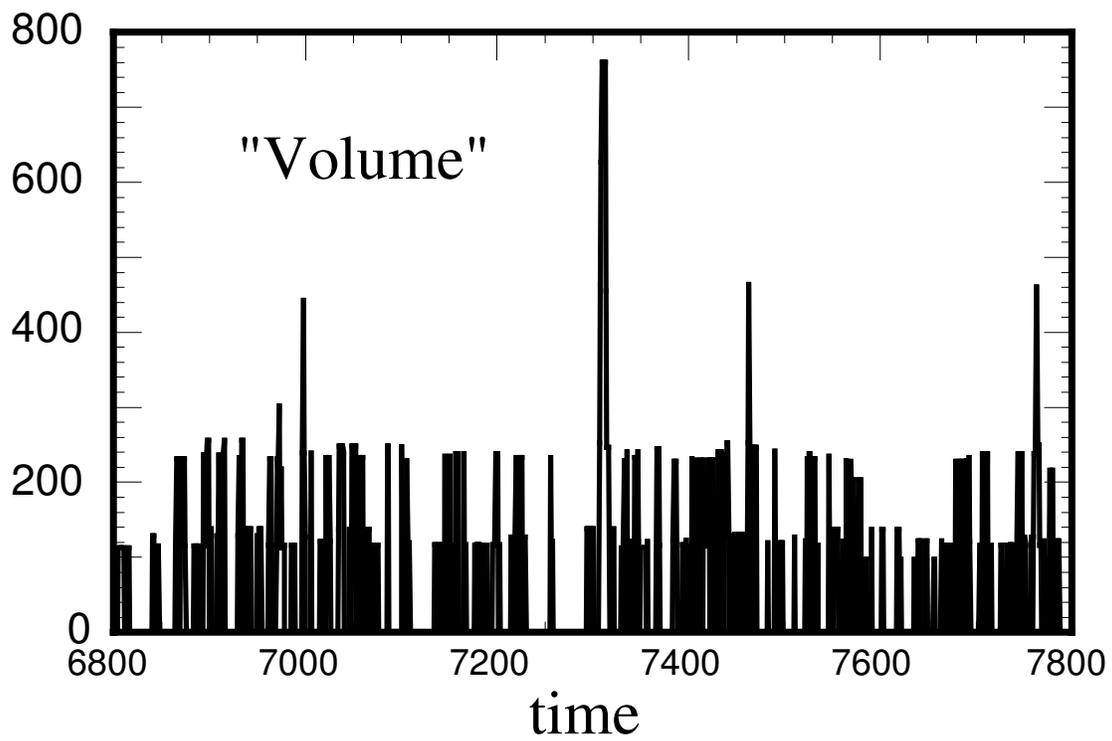
\end{document}